\documentclass[aps,prl,twocolumn]{revtex4-1}

\usepackage{graphicx}
\usepackage{epstopdf}
\usepackage{color}
\usepackage{epsfig}
\usepackage[english]{babel}
\usepackage{amsmath}
\usepackage{amssymb}
\usepackage{natbib}

\newcommand{\LGF}{$\rm LiGdF_4$}

\begin{document}

\title{Strongly anisotropic magnetocaloric effect in a dipolar magnet LiGdF$_4$}

\author{G.\,Iu.\,Andreev}
\affiliation{Kazan Federal University, 420008 Kazan, Russia}
\author{I.\,V.\,Romanova}
\affiliation{Kazan Federal University, 420008 Kazan, Russia}
\author{O.\,A.\,Morozov}
\affiliation{Kazan Federal University, 420008 Kazan, Russia}
\affiliation{Zavoisky Physical-Technical Institute, FRC Kazan Scientific Center of RAS,  420029 Kazan, Russia}
\author{S.\,L.\,Korableva}
\affiliation{Kazan Federal University, 420008 Kazan, Russia}
\author{R.\,G.\,Batulin}
\affiliation{Kazan Federal University, 420008 Kazan, Russia}
\author{V.\,N.\,Glazkov}
\affiliation{P. Kapitza Institute for physical problems RAS, 117334 Moscow, Russia}
\affiliation{HSE University, 101000 Moscow, Russia}
\author{S.\,S.\,Sosin}
\email{sosin@kapitza.ras.ru}
\affiliation{P. Kapitza Institute for physical problems RAS, 117334 Moscow, Russia}
\affiliation{HSE University, 101000 Moscow, Russia}

\date{\today}

\begin{abstract}
We report the detailed study of the magnetocaloric effect (MCE) in a dipolar-Heisenberg magnet \LGF\ using magnetization measurements performed on a single crystal sample. Entropy variation on isothermal demagnetization from the magnetic field up to 3~T is determined in the temperature range 2--10~K for two principal directions of the applied field (parallel and perpendicular to the tetragonal $c$-axis of the crystal). The MCE is found to be highly anisotropic, with the cooling efficiency being up to twice higher at $H\parallel c$. The results are nicely interpreted in the frame of a conventional molecular field approach taking into account considerable anisotropy of the paramagnetic Curie-Weiss temperature. These results are compared to earlier studies of MCE in powder samples of \LGF\ [T.~Numazawa {\it et al}., AIP Conf. Proc. \textbf{850}, 1579 (2006)] as well as with analogous data for other well known magnetocaloric materials. Our findings may open new possibilities to enhance the efficiency of magnetic refrigeration in the liquid helium-4 temperature range.
\end{abstract}

\pacs 75.30.Sg, 75.20.-g, 75.40.Cx

\maketitle

\section{Introduction}
Adiabatic demagnetization is an efficient tool to achieve low temperatures, being one of the beautiful manifestations of the basic thermodynamics principles: the release of magnetic entropy on lowering the magnetic field leads to the decrease of the lattice entropy, thus resulting in cooling of the entire sample. Demagnetization of paramagnetic salts originally discussed by P.~Debye~\cite{Debye}, was the first instrument to reach sub-kelvin temperatures. Nuclear demagnetization stage~\cite{Kurti} till now remains the only route to achieve sub-millikelvin range in dilution fridge cryostats. Conventional methods of magnetic cooling for practical applications are now well developed and reviewed in textbooks (see {\it e.g.}~\cite{Lounasmaa}). From the practical point of view magnetic refrigerators have considerable convenience of simple and compact construction as well as its independence of gravity and avoiding expensive and sparse $^3$He-based cooling agent. However, particular disadvantage of paramagnetic salts is the small concentrations of magnetic ions in the substance resulting in low entropy density and insufficient cooling capacity of these materials.

An attempt to increase the entropy density in regular paramagnets generally encounters the problem of growing magnetic interactions, for example dipolar, and thus, limiting the temperature range of efficient cooling. The later can be partly avoided in systems with competing magnetic interactions. The family of  magnets with strongly frustrated principal exchange interaction is believed to be promising in this respect. An ``infinite'' degeneracy of the ground state and a macroscopic number of soft modes in the excitation spectrum leaves a finite part of entropy of a concentrated system unfrozen at a temperature scale much lower than the energy of the principal interaction~\cite{gardner_review}. Lifting this degeneracy by magnetic field in a spin-saturated state opens broad space for an enhanced magnetocaloric effect~\cite{mzh03} in various ranges of temperatures and magnetic fields, as was observed for some types of rare-earth garnets (see {\it e.g.}~\cite{Numazawa_DGGG} or more recent research~\cite{Bras}) or pyrochlores~\cite{sosin,Wolf}.

In the present paper we discuss a fresh look at magnetocaloric properties of a lithium-gadolinium fluoride \LGF. The absence of magnetic ordering down to at least 400~mK in combination with exceptionally high entropy density have already attracted much attention to this system as one of promising powder magnetic refrigerants~\cite{Numazawa06,Numazawa09,Wikus14}. Lack of microscopic model underlying the magnetic disorder was filled by recent experiments performed on single-crystal samples of a concentrated \LGF\ and strongly diluted $\rm LiY_{1-x}Gd_xF_4$ which reveal an unusual type of ``hidden'' magnetic frustration, {\it i.e.} a competition between {\it various} types of interactions~\cite{sosin1}. Moreover, the fine compensation of contributions from exchange coupling, long-range dipolar interaction and single-ion anisotropy to the magnetic susceptibility makes the Curie-Weiss temperature strongly anisotropic, being very close to zero for one of the principal directions of the external field. Here we demonstrate that applying the magnetic field along the tetragonal axis of a single-crystal sample makes the demagnetization process up to 30\% more efficient than that previously observed for a powder material, thus opening new ways to enhance the magnetic refrigeration at liquid helium-4 temperature range under moderate applied fields.

\section{Experimental results}

\begin{figure}
\centering
\includegraphics[width=0.9\columnwidth]{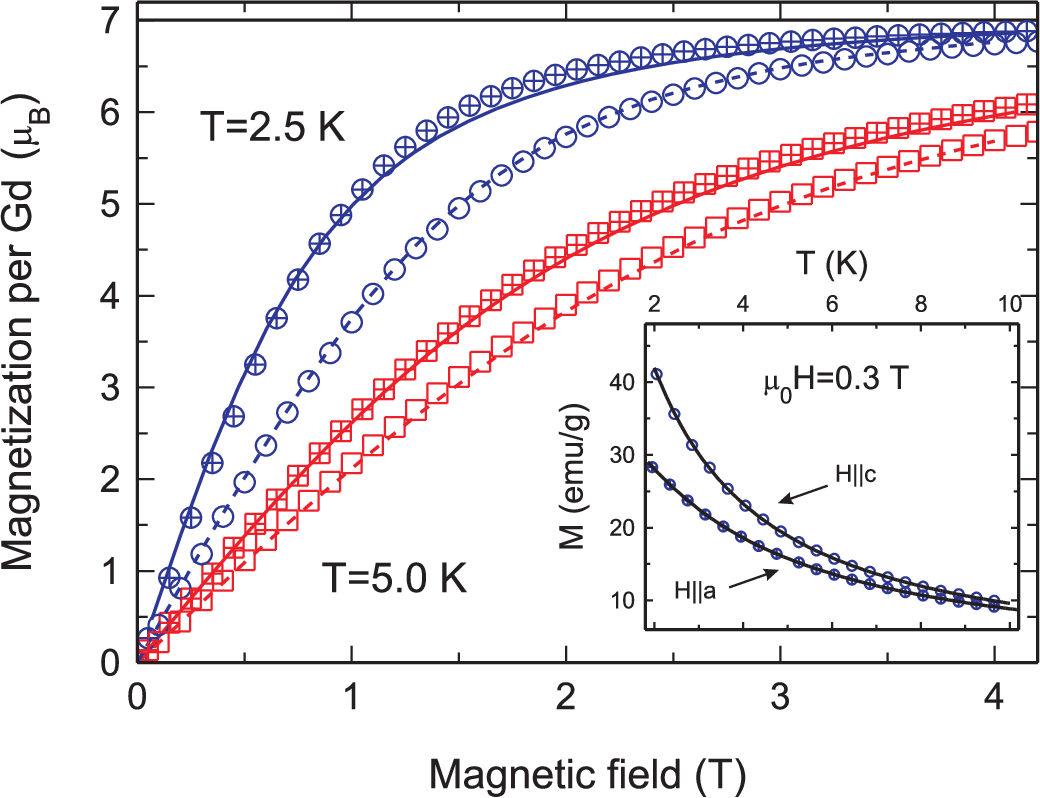}
\caption{Magnetization in \LGF\ measured in magnetic field $H\parallel c$ ($\oplus$ and $\boxplus$) and $a$-axes ({\large $\circ$} and $\Box$) at temperatures $T=2.5$ and 5.0~K. Solid and dashed lines show the results of mean-field calculations~(\protect\ref{brillouin}) with $\theta_{\rm cw}^c=0$ (an ideal paramagnet) and $\theta_{\rm cw}^a=-1.4$~K, respectively. Inset: typical magnetization vs temperature curves (every 4th point shown) measured under constant field ($\mu_0H=0.3$~T) applied along to principal axes, $c$ and $a$; solid lines are polynomial fits used for computing derivatives. The whole set of data consists of sixty pairs of curves recorded in the field range 0.05 to 3~T with the step of 0.05~T.}
\label{magn-susc}
\end{figure}

The crystal structure of \LGF\ is of Scheelite-type with the space group $I4_1/a$ (C$_{\rm 4h}^6$) and the local symmetry S$_4$ on each Gd-site. The tetragonal unit cell with the parameters $a=5.219$ and $c=10.97$~{\AA} contains four formula units~\cite{Keller}. A single-crystal sample was grown using a standard Bridgman-Stockbarger technique. The directions of crystal axes were precisely determined by X-ray Laue diffraction patterns.

Magnetization measurements have been carried out using the Quantum Design PPMS Vibrating Sample Magnetometer. The sample was cut from the parent single crystal in a shape of a thin plate 16.8~mg by mass containing the $ac$ crystal plane. A magnetic field $H$ has been applied along the two principal crystal axes, $c$ and $a$, within the sample plane to exclude the demagnetization corrections. The temperature of the experiment varied from 2 to 10~K with the data obtained on cooling and heating being indistinguishable. The isothermal magnetization curves recorded at $T=2.5$ and 5~K for two directions of the external field $H\parallel c,a$ are presented in the main panel of Fig.~\ref{magn-susc}. Linear low-field parts of the curves demonstrate significant anisotropy of the susceptibility amounting to $\chi^c/\chi^a\simeq 1.5$ at $T=2.5$~K. This anisotropy, clearly visible by  temperature dependences of magnetization measured at small constant field (see Inset), reflects the above mentioned anisotropic paramagnetic Curie-Weiss temperature.

\begin{figure}
\centering
\includegraphics[width=0.93\columnwidth]{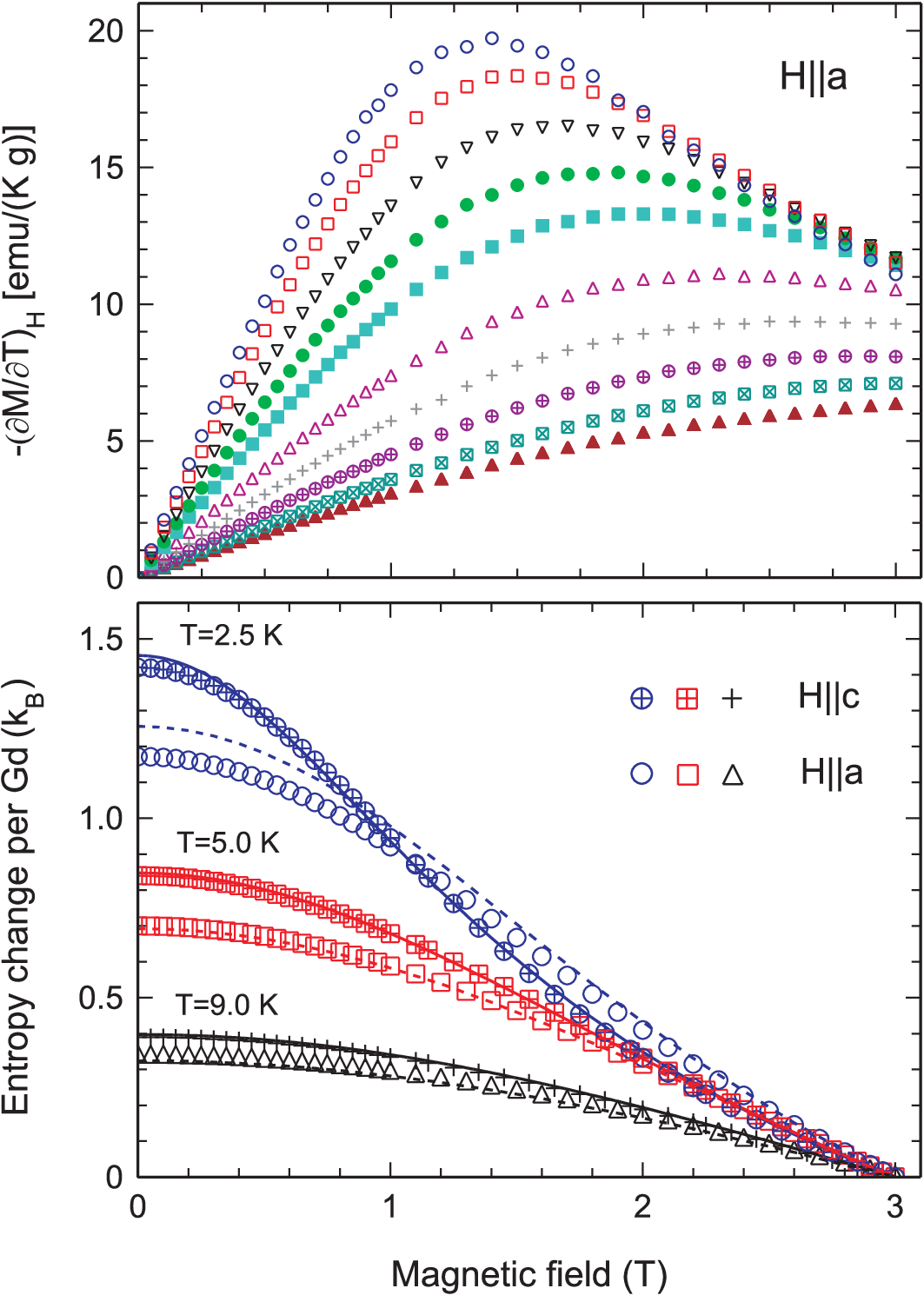}
\caption{Upper panel: field dependence of the derivatives of $M(T)$ curves computed at given temperatures (from top to bottom: 2.1, 2.5, 3.0, 3.5, 4.0, 5.0, 6.0, 7.0, 8.0, 9.0~K); here the magnetic field is applied along $a$-axis (analogous procedure was also carried out for curves obtained at $H\parallel c$). Lower panel: Field dependence of the magnetic entropy change on isothermal decreasing the starting field from $\mu_0H_{\rm i}=3$~T to zero at various temperatures. Symbols are explained in the legend, solid and dashed lines are drawn using formula~(\protect\ref{entropy}) with $\theta_{\rm cw}^c=0$ (an ideal paramagnet) and $\theta_{\rm cw}^a=-1.4$~K, respectively.}
\label{dMdT-dS}
\end{figure}

The standard procedure to study the MCE from static magnetization data involves collecting a set of $M(T)$ curves measured at constant fields (from 0.05 to 3~T with a step of 0.05~T in our experiment). Using the Maxwell relation $(\partial M/\partial T)_H=(\partial S/\partial H)_T$ one obtains a set of isothermal field dependences of the derivative $(\partial S/\partial H)_T$ (see upper panel of Fig.~\ref{dMdT-dS}) which can be then integrated over the field to obtain the entropy change for the demagnetization process performed at various temperatures. A few examples of these curves integrated at several temperatures for two principal directions of the applied field are shown in the lower panel of Fig.~\ref{dMdT-dS}.

\begin{figure}
\centering
\includegraphics[width=0.92\columnwidth]{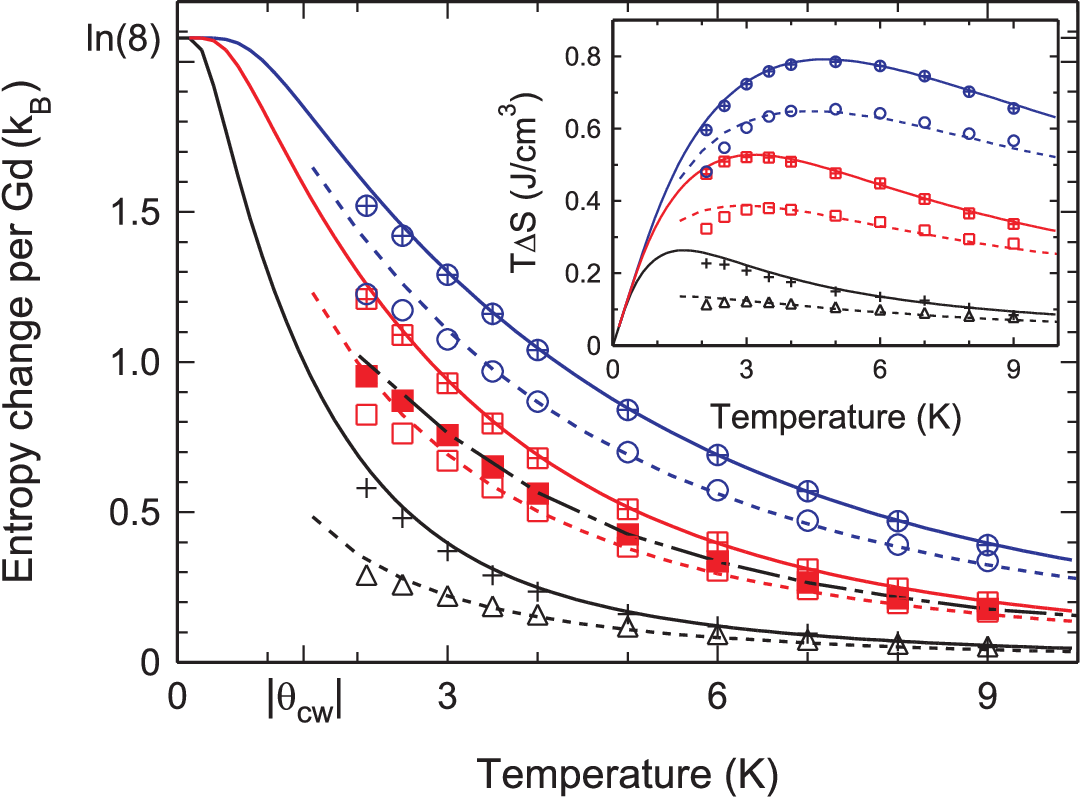}
\caption{Magnetic entropy change at an isothermal field decrease from $\mu_0H_{\rm i}=3$, 2 and 1~T to zero (circles, squares and triangles-crosses respectively) performed at various temperatures; $\oplus$, $\boxplus$ and $+$ symbols correspond to the field direction $H\parallel c$, open symbols are for $H\parallel a$-axis, $\blacksquare$ is the powder average of data for $\mu_0H_{\rm i}=2$~T ($\boxplus$ and $\square$) as described in the text. Solid and dashed lines are calculated using formula~(\protect\ref{entropy}) with $\theta_{\rm cw}^c=0$ and $\theta_{\rm cw}^a=-1.4$~K, respectively, dashed-dotted line represents data from Ref.~\protect\cite{Numazawa06}. The inset shows the corresponding specific cooling capacities of the demagnetization process performed at constant temperatures.}
\label{entropy-cooling}
\end{figure}

Final results, that are magnetic entropy change on the isothermal demagnetization from magnetic fields $\mu_0H_{\rm i}=3,~2$ and 1~T applied along to principal axes $c$ and $a$, are presented in the main panel of Fig.~\ref{entropy-cooling}. The demagnetization process traced at various temperatures reveals considerable anisotropy, being up to twice more efficient under field $H\parallel c$, when the Curie-Weiss temperature $\theta_{\rm cw}^c\simeq 0$ and the system is expected to be very close to an ideal paramagnet.

\section{Discussion and concluding remarks}
The above experimental results can be adequately interpreted in terms of the standard approach to a system of interacting magnetic moments in a disordered state (see {\it e.g.}~\cite{smart}). The interaction is introduced in the form of the effective molecular field acting on a single moment from the rest of the ensemble, which in the antiferromagnetic case can be written as $H_M=-\lambda M$, where $\lambda>0$ is a molecular field constant, $M$ is a uniform net magnetization per spin. This molecular field leads to a non-zero Curie-Weiss temperature in the Curie-like paramagnetic susceptibility, which can be directly related to the molecular field strength: $\theta_{\rm cw}=-C\lambda$, where $C=\frac{(g\mu_{\rm B})^2S(S+1)}{3k_{\rm B}}$ is a Curie constant ($g\simeq 2.0$ is a $g$-factor of a Gd$^{3+}$ ion, $\mu_{\rm B}$ is a Bohr magneton, $k_{\rm B}$ is a Boltzmann constant). For arbitrary $(H,T)$ values the magnetization can be self-consistently expressed in the form of a Brillouin function with the external field $H$ replaced by a sum of external and molecular fields:
\begin{equation}
{\tilde M}=B_{7/2} (H-{\tilde\lambda}{\tilde M}),
\label{brillouin}
\end{equation}
where ${\tilde M=M/(g\mu_{\rm B}S)}$ is a reduced magnetization per spin and ${\tilde \lambda}=g\mu_{\rm B}S\lambda\simeq 7.0$~kOe (for $\theta_{\rm cw}=-1.4$~K) is a molecular field constant expressed in CGS field units. The value of $M$ obtained from~(\ref{brillouin}) can be used to find the magnetic entropy per moment as a function of field and temperature in the following form~\cite{smart}:
\begin{equation}
T{\cal S}={\cal E}-{\cal F}=\left (\frac{\lambda}{2}M^2-MH\right )+k_{\rm B}T\ln {\left [\frac{\sinh{\frac{(2S+1)}{2}x}}{\sinh{\frac{x}{2}}} \right ]},
\label{entropy}
\end{equation}
where ${\cal E}$ and ${\cal F}$ are an energy and free energy per magnetic moment, respectively, $x=g\mu_{\rm B}(H-\lambda M)/(k_{\rm B}T)$. Theoretical curves for $\theta_{\rm cw}^c=0$ (an ideal paramagnet) and $\theta_{\rm cw}^a=-1.4$~K are shown in Figs.~\ref{magn-susc}--\ref{entropy-cooling} by solid and dashed lines, respectively.

\begin{table}
\caption{Comparison of potential magnetocaloric cooling capacities for different compounds. Entropy changes expressed in $k_{\rm B}$ per magnetic ion on demagnetization from the initial field $\mu_0H_{\rm i}=1$ and 2~T to zero at constant temperature $T=2$~K are given in 3d and 4th columns respectively.}
\centering
  \begin{tabular}{|c|c|c|c|c|}
  \hline
  &Magnetic ions&\multicolumn{2}{|c|}{Entropy change}&\\
  Material&density&\multicolumn{2}{|c|}{from $\mu_0H_{\rm i}$}& Ref.\\
  \cline{3-4}
   & $(10^{22}/{\rm cm}^{3})$ & ~~~1~T~~~ & ~~~2~T~~~ &\\ \hline
  Gd$_3$Ga$_5$O$_{12}$ & 1.26 & 0.11 & 0.72 & \protect\cite{Numazawa_DGGG} \\ \hline
  Dy$_3$Ga$_5$O$_{12}$ & 1.24 & 0.45 & 0.65 & \protect\cite{Numazawa_DGGG} \\ \hline
  KBaYb(BO$_3$)$_2$ & 0.68 & 0.11 & 0.31 & \protect\cite{Tokiwa}\\ \hline
  \LGF\ (powder) &  & -- & 1.02 & \protect\cite{Numazawa06} \\ \cline{1-1} \cline{3-5}
  \LGF\ $H\parallel c$ & 1.34 & 0.63(2) & 1.23(2) & This \\ \cline{1-1} \cline{3-4}
  \LGF\ $H\parallel a$ &   & 0.30(2) & 0.84(2) & work \\ \hline
  \end{tabular}
\label{table}
\end{table}

One can see that theoretical curves computed for an ideal paramagnet perfectly reproduce our data for $H\parallel c$ in the whole temperature range 2--10~K. As was mentioned above, all substantial magnetic interactions (exchange, dipolar and single-ion anisotropy) contributing to the susceptibility compensate each other in the way that the system is magnetized along $c$-axis in a wide temperature range effectively as non-interacting magnetic ions. However, when the magnetic field is applied in a perpendicular direction, the compensation is broken and magnetic ions are subjected to a relatively strong internal molecular field. The results of a molecular field approximation with an antiferromagnetic Curie-Weiss temperature $\theta_{\rm cw}=-1.4$~K are also in a good agreement with the experimental data obtained for $H\parallel a$. This anisotropy immediately leads to a considerable difference in the amount of entropy released on demagnetizing in the two principal directions of the external field. The ratio $\Delta S_{c}/\Delta S_{a}$ can reach a factor of 2 at $T=2$~K for the starting field $\mu_0H_{\rm i}=1$~T (Fig.~\ref{entropy-cooling}). One should note that analogous results measured in a powder sample~\cite{Numazawa06} and shown in this Figure by bold dashed-dotted line appear to fall between our data obtained for $H\parallel c$ and $a$-axes. Moreover, averaging our results by orientations in the powder as $\Delta S_{\rm p}=1/3\Delta S_{c}+2/3\Delta S_{a}$ one achieves perfect quantitative agreement with the previous data.

Further, we have summarized in Table~\ref{table} the entropy changes (expressed in $k_{\rm B}$ per magnetic ion) on demagnetizing from two different starting fields $\mu_0H_{\rm i}=1$ and 2~T measured in some other state-of-art cooling materials as well as in the powder \LGF\ known from literature. The comparison both with two well known rare-earth gallium garnets~\cite{Numazawa_DGGG} and with KBaYb(BO$_3$)$_2$, a recently studied frustrated material suitable for cooling to very low temperatures~\cite{Sanders,Tokiwa}, is obviously greatly in favor of a single-crystal \LGF\ demagnetized at $H\parallel c$. The advantage of \LGF\ is especially pronounced if one takes into account the enhanced density of magnetic ions in this material which is important from the practical point of view. Our data (see inset to Fig.~\ref{entropy-cooling}) show that the most efficient demagnetization process in \LGF\ could be achieved under moderate applied fields. In the temperature range 3--4~K the highest cooling efficiency is observed for a starting magnetic field $\mu_0H_{\rm i}\simeq 2$~T while that for 1--2 K shifts to smaller initial fields around 1~T. In both regimes the cooling capacity of \LGF\ reaches the value $\simeq 0.25$~J/T per cm$^3$ of the material which enables the cooling power up to $\simeq 10$~mW/cm$^3$ for a reasonable field sweep rate 2T/min accessible in typical laboratory cryomagnets.

To summarize, using static magnetization measurements of a single-crystal \LGF\ performed in a temperature range 2--10~K we have demonstrated the MCE in the system to be considerably anisotropic. This anisotropy results from competing contributions from various magnetic interactions to the paramagnetic susceptibility of the system. We show that when the magnetic field is applied along the tetragonal axis of the crystal, \LGF\ is magnetized in a wide temperature range in the way similar to a system of non-interacting magnetic moments, thus enhancing the MCE to a maximum possible level of an ideal paramagnet. These results can be described in the frame of a usual molecular field approach taking into account considerable anisotropy of the paramagnetic Curie-Weiss temperature. Comparison with other well known materials for magnetic refrigeration shows significant advantage of a single crystal \LGF\ for demagnetization at temperatures of liquid helium-4 (1--4~K) in moderate applied fields, which may open new opportunities for practical applications.

\section{Acknowledgments}

The work was financially supported by: Russian Science Foundation, Grant No 22-12-00259 (sample growth); Basic research program of HSE University (data processing and theoretical calculations); Kazan Federal University Strategic Academic Leadership Program PRIORITY-2030 (magnetization measurements).

\end{document}